# A Case for AI Safety via Law

**Jeffrey W. Johnston**
jeffj4a@earthlink.net

July 31, 2023

## Abstract

How to make artificial intelligence (AI) systems safe and aligned with human values is an open research question. Proposed solutions tend toward relying on human intervention in uncertain situations, learning human values and intentions through training or observation, providing off-switches, implementing isolation or simulation environments, or extrapolating what people would want if they had more knowledge and more time to think. Law-based approaches—such as inspired by Isaac Asimov—have not been well regarded. This paper makes a case that *effective legal systems are the best way to address AI safety*. Law is defined as any rules that codify prohibitions and prescriptions applicable to particular agents in specified domains/contexts and includes processes for enacting, managing, enforcing, and litigating such rules.

## 1  Question Presented

Whether laws and legal processes are an effective way to make AIs safe and aligned with human values.

## 2  Statement of the Case

Providing mechanisms for making AI systems safe and aligned with human values is increasingly important as AI technology advances and its risks become more salient (Wikipedia, AI safety; Wikipedia, AI alignment; FLI, 2023; Yudkowsky, 2023). The need for safe and aligned AIs applies to systems that are narrowly focused (weak, narrow, GOFAI), general purpose (strong, AGI), and superintelligent (ASI) (Wikipedia, Weak artificial intelligence; Wikipedia, Artificial general intelligence; Wikipedia, Superintelligence; Carlsmith, 2023).

AI safety and alignment risks include:

- Concerns often directed at narrow AI such as automation-spurred job loss, privacy threats, deepfake proliferation, algorithmic bias, socioeconomic inequality, market volatility, and weapons automatization (Thomas, 2023),
- Malicious use of AI by humans such as leveraging AI for mass destruction (e.g., bioweapons) or perpetrating other harms at scale (e.g., cyberattacks, financial fraud, propagandizing) (Brundage et al., 2018),
- Humans becoming overly dependent on AI (Anderson et al., 2018),
- *Outer misalignment / specification gaming*[1] where AIs misinterpret human-specified goals (or exploit bugs or loopholes) to harmful effect. Canonical examples include King Midas turning everything he touches to gold and runaway paperclip optimizers, and
- *Inner misalignment / instrumental convergence to dangerous values* where AIs (esp. AGIs or ASIs) become inclined to self-preservation, resisting change to original goals, enhancing their cognitive abilities, developing better technologies, and acquiring resources[2].

Proposed solutions tend toward relying on human intervention in uncertain situations (Pynadath and Tambe, 2001); having AIs learn human values, intentions, or preferences via observation, training, feedback, or debate (Riedl and Harrison, 2016; Russell, 2017; Christiano et al., 2017; Noothigattu et al., 2017; Soares, 2018; Irving et al., 2018; Russell, 2019); providing effective off-switches (Orseau and Armstrong, 2016); utilizing verified, isolated, or simulated environments (Arnold and Scheutz, 2018); designing AIs to be communicative, corrigible, and transparent with human collaborators (Soares et al., 2015; Briggs and Scheutz, 2015; Christiano, 2017); avoiding specification and design errors (Amodei et al., 2016); transferring control to the most competent agent (Pynadath and Tambe, 2001; Scerri et al., 2002); or trying to have AIs extrapolate what people really want if they had more knowledge and time to think (Yudkowsky, 2004; Tarleton, 2010)[3]. Russell et al. (2016) provide an overview of short-term and long-term AI research priorities, and consider AI safety ("robustness") in terms of verification, validity, security, and control. Bostrom (2014, Table 10) characterizes approaches "for dealing with the agency control problem at the heart of AI safety" as: boxing methods, incentive methods, stunting, tripwires, direct specification, domesticity, indirect normativity, and augmentation. Krakovna (2022) provides references to recent AI alignment proposals.

---

[1] "Specification gaming is a behaviour that satisfies the literal specification of an objective without achieving the intended outcome" (Krakovna et al., 2020). See (Krakovna, 2023) for a table of such behaviors that have been observed in AI systems from 1983 to 2023. A classic example is when Lenat's "Eurisko won the Trillion Credit Squadron (TCS) competition two years in a row creating fleets that exploited loopholes in the game's rules, e.g. by spending the trillion credits on creating a very large number of stationary and defenseless ships."

[2] Bostrom (2014) articulated the *instrumental convergence thesis* suggesting that certain values will arise "in sufficiently advanced AI systems" since they would be useful (instrumental) for achieving a wide variety of other goals. These values include *self-preservation, resisting changes to original goals, enhancing cognitive abilities, developing better technologies, and acquiring resources*. The thesis was based on Omohundro (2009) who postulated "a number of 'drives' that will appear in sufficiently advanced AI systems of any design." Others subscribing to the thesis include Russell (2017) who focuses on AI *self-preservation*, Yudkowsky (2022, item -3) who endorses *orthogonality* and *instrumental convergence*, and Carlsmith (2023) who characterizes the concern as *power-seeking behavior*. See also (Wikipedia, Instrumental convergence).

Law-based approaches, such as inspired by Isaac Asimov (1942)[4], have not been well regarded. Yampolskiy (2013) writes: "The general consensus seems to be that no set of rules can ever capture every possible situation and that interaction of rules may lead to unforeseen circumstances and undetectable loopholes leading to devastating consequences for the humanity." Many cite Asimov's Laws and stories as demonstrations of how laws fail and cannot effectively direct AI behaviors (Dvorsky, 2014; Yudkowsky, 2004, p. 16; Yudkowsky, 2016; Bench-Capon and Modgil, 2016, p. 2; Awad et al., 2018, p. 59; Kuipers, 2018, p. 91). Legal systems are justifiably criticized for being flawed or dysfunctional—plagued by inconsistencies; imprecision; bias; jurisdictional conflicts; corruption; subjectivity; excessive complexity; high cost; susceptibility to gaming; tendency to fail in novel situations; slowness to enforce, adjudicate, and amend; and other issues.

However, precedents supporting law-oriented approaches for AI safety exist and include:

1. Asimov (1981) who argues three laws are fundamental ("obvious from the start") for assuring *every tool* a human uses is safe, effective, and durable, whether it is a robot, a knife, or the US Constitution.
2. Weld and Etzioni (1994) propose rule primitives that might be used in AI planning frameworks for constraining robot behavior, i.e., *dont-disturb* and *restore*.
3. Rissland et al. (2003) discuss the nature of law, *AI and Law* history, and the field circa 2003—including work on automating legal reasoning.

---

[3] Yudkowsky (2004) calls this Coherent Extrapolated Volition (CEV) and defines it poetically as "our wish if we knew more, thought faster, were more the people we wished we were, had grown up farther together; where the extrapolation converges rather than diverges, where our wishes cohere rather than interfere; extrapolated as we wish that extrapolated, interpreted as we wish that interpreted."

[4] See next section for text of Asimov's Laws.

4. Omohundo (2008), who convinced many in the AI field that dangerous instrumental goals ("drives") would spontaneously emerge in powerful AIs, recommended "we should begin by designing a 'universal constitution' that identifies the most essential rights we desire for individuals and creates social mechanisms for ensuring them in the presence of intelligent entities of widely varying structures." (Omohundro, 2013) concludes: "It appears that humanity's great challenge for this century is to extend cooperative human values and institutions to autonomous technology for the greater good."
5. Johnston (2009) proposes, "Why not require all AGIs be linked to a single large database of law—legislation, orders, case law, pending decisions—to account for the constant shifts [in legal definitions, interpretations, social context, and political acceptability]? Such a corpus would be ever changing and reflect up-to-the-minute legislation and decisions on all matters man and machine. Presumably there would be some high level guiding laws, like the US Constitution and Bill of Rights [… And, when necessary, an AGI would] inform its action using analysis of the deeper corpus. Surely a 200-volume set of international law would be a cakewalk for an AGI. The latest version of the corpus could be stored locally in most AGIs and just key parts local in low end models—with all being promptly and wirelessly updated as appropriate. This seems like a reasonable solution given the need to navigate in a complex, ever changing, context-dependent universe."
6. Hanson (2009) writes about law-abiding robots where, "In the long run, what matters most is that we all share a mutually acceptable law to keep the peace among us, and allow mutually advantageous relations, not that we agree on the 'right' values."
7. Genesereth (2015) discusses *Computational/Embedded Law*, noting how it can make humans (and presumably AIs) "aware of the legal status of our actions as we are performing them." He offers a metaphor of *The Cop in the Backseat*: "a friendly policeman in the backseat of our car […], real or computerized, [that] could offer regulatory advice as we drive around—telling us speed limits, which roads are one-way, where U-turns are legal and illegal, where and when we can park, and so forth."
8. Genesereth (2016) introduces *Corpus Legis*—"a library of governmental regulations encoded in computable form."
9. Prakken (2016) suggests "the current fruits of AI & law research on supporting human legal decision making can be used for making autonomous artificial systems behave lawfully" (although current approaches are inadequate).

10. Wolfram (2016) notes that Gottfried Liebniz's dream of "turning human law into an exercise in computation […] didn't succeed. But three centuries later, […] we're finally ready to give it a serious try again and […] it's likely to be critical to the future of our civilization and its interaction with artificial intelligence." He suggests his Wolfram Language (a *symbolic discourse language*) might eventually be usable for *computational contracts* and for providing codes of conduct "that AIs can readily make use of."
11. Etzioni (2017) proposes three rules that may be particularly effective in steering AI. (See Argument for details.)
12. Kuipers (2018) focuses on "the key role of trust in human society" and how social norms are used to promote trust. He defines social norms to include "morality, ethics, and convention, *sometimes encoded and enforced as laws*, sometimes as expectations with less formal enforcement." He asserts, "intelligent robots […] must be able to understand and follow social norms" and such an ability may be implemented using a hybrid ethics architecture (combining virtue ethics, deontology, and utilitarianism) that enables "fast but fallible pattern-directed responses; slower deliberative analysis of the results of previous decisions; and, yet slower individual and collective learning processes."
13. O'Keefe (2022) argues that "working to ensure that AI systems follow laws is a worthwhile way to improve the long-term future of AI." He proposes "what an ideal law-following AI (LFAI) system might look like."
14. Bai et al.'s (2022) *Constitutional AI* approach demonstrates an ability to "train less harmful [Large Language Model] systems entirely through the specification of a short list of principles or instructions, i.e., a constitution."
15. Nay's (2023) *Law Informs Code* proposal contends, "The target of AI alignment should be democratically endorsed law," "Data generated by legal processes and the tools of law (methods of law-making, statutory interpretation, contract drafting, applications of standards, and legal reasoning) can facilitate the robust specification of inherently vague human goals to increase human-AI alignment," "If properly parsed, [law] distillation offers the most legitimate computational comprehension of cosocietal values available," and "[Law is] the applied philosophy of multi-agent alignment." Nay also argues: (1) Legal Theory is Well-Developed and Applicable to Alignment, (2) Legal Informatics Can Scale with AI Capabilities, and (3) Legal Processes, Data & Experts Can Improve AI.

Some empirical support for the effectiveness of non-law-based approaches can be found in the papers cited in the second paragraph of this section, but that evidence is generally preliminary and weak. Evidence for the effectiveness of law-based approaches is more compelling. Law "is closely connected to the development of civilizations" (Wikipedia, Legal history) and has served as a stabilizing force for societies of intelligent agents throughout history. Allot (1981, p. 229) argues that, "without a generally respected and effective legal system, a society will tend to its own disintegration." The World Justice Project claims, "[Law] is the foundation for communities of justice, opportunity, and peace—underpinning development, accountable government, and respect for fundamental rights. Research shows that rule of law correlates to higher economic growth, greater peace, less inequality, improved health outcomes, and more education" (WJP, 2019).

## 3 Argument

*Effective legal systems are the best way to address AI safety*. We substantially agree with the above-quoted claims of Nay (2023). The approach argued herein is called *AISVL* (for *AI Safety Via Law*) to distinguish it from similar proposals.

Many of the proposed *non*-law-based solutions may be worth pursuing to help assure AI systems are law abiding. However, they are secondary to having a robust, well-managed, readily available corpus of codified law—and complimentary legal systems—as the foundation and ultimate arbiter of acceptable behaviors for all intelligent systems, both biological and mechanical.

*To have safe, aligned, viable societies, AIs and humans must know the law, strive to abide by it, and be subject to effective intervention when violated*. These three requirements apply to humans in most modern societies and are generally—if imperfectly—achieved through legal systems. They should apply to all intelligent agents and systems.

AISVL recognizes that a small set of static rules (e.g., Asimov's Laws, The Golden Rule) or value-preserving utility functions are not feasible. Rather, *intelligent systems must comply with a large and dynamic set of laws that are drafted, enacted, enforced, litigated, and maintained over time via full-featured jurisprudence systems*.

AISVL defines laws broadly as *any rules that codify prohibitions and prescriptions applicable to particular agents in particular domains/contexts and are sufficiently binding*. Codification requires that the rules be maintained in authoritative repositories that can be accessed and interpreted by everyone. To be sufficiently binding, effective motivations must exist for agents to comply—motivations such as provided through education, social pressure, coercion, and/or other enforcement mechanisms. All intelligent systems should have ready access to the latest versions of laws relevant to their operational contexts. Whereas AIs and humans are generally black boxes, Law, critically, is a white box.

Laws by the above definition include: constitutions, statutes (legislation), decrees, executive orders, regulations, court decisions (case law), treaties, contracts (e.g., sales agreements, service agreements, leases, EULAs, warranties, NDAs), rules (defined by governments, homeowner associations, households, classrooms, businesses, associations, and other organizations), best practices, policies, codes of conduct, standards, principles, and similar rules. Legal domains range from local, regional, national, and international governance (classic social contracts, a.k.a. public law) to private contracts, rules, laws, and norms applicable in all kinds of economic and social interactions between agents and institutions—from basic principles adopted by organizations to games and sports to product and service agreements and more.

AISVL recognizes the *essential equivalence and intimate link between democratically developed law* (Nay, 2023, p. 11) *and consensus ethics. Both are human inventions intended to facilitate the wellbeing of individuals and the collective.* They represent shared values culturally determined through rational consideration and negotiation. To be effective, democratic law and consensus ethics should reflect sufficient agreement of a significant majority of those affected. Democratic law and consensus ethics *are not* inviolate physical laws, instinctive truths, or commandments from deities, kings, or autocrats. They *do not* represent individual values, which vary from person to person and are often based on emotion, irrational ideologies, confusion, or psychopathy.

Bodies of law have historically been based on ethical values, which are used to inform the wider body. These values are often stated in introductions of foundational documents. For example, key values called out in the preamble of the US Constitution (1787) are unity, justice, domestic tranquility, common defense, general welfare, and liberty (http://constitutionus.com). Key values expressed in the US Declaration of Independence (1776) are equality[5] and rights to life, liberty, and the pursuit of happiness (Wikipedia, United States Declaration of Independence). In ancient Egyptian Law (3000 BCE), the values of tradition, rhetorical speech, equality, and impartiality were central (Wikipedia, Legal history). In the Code of Ur-Nammu (2100 BCE), truth and equity are prominent (Wikipedia, Code of Ur-Nammu). The Code of Hammurabi (1754 BCE) promoted values of justice, destruction of the wicked and evil, preventing the strong from harming the weak, subjugating the "Black Head Race," enlightening the land, and furthering the welfare of mankind (Wikipedia, Code of Hammurabi). The Universal Declaration of Human Rights (UN, 1948) focused on values of freedom (of movement, thought, conscience, speech, religion, peaceful assembly, marriage, community participation), equality, liberty, security, humane treatment, access to legal remedies (public hearings, presumed innocence), asylum from prosecution, right to own property, government by will of the people, safe and equitable employment, right to rest and leisure, social security (in health, well-being, education), intellectual property protection, and duties to the community for all people.

---

[5] Equality and justice have only recently begun to apply to all people in modern liberal societies and not just members of privileged groups.

Accordingly, legal systems consist of a core set of moral values (a virtue/deontological core) surrounded by a large corpus of legal refinements (a consequentialist/utilitarian shell),[6] where multiple systems coexist and apply per different jurisdictions and subject areas, and change over time. Core values reflect the spirit of the law. Consequentialist shells specify its letter. This nexus of democratic law and consensus ethics provides a solid foundation for AI safety and value alignment.

To operationalize AISVL, *societies would begin by adopting existing bodies of law* with additions like Etzioni's (2017) rules:

1. An AI system must be subject to the full gamut of laws that apply to its human operator[7],
2. An AI system must clearly disclose that it is not human, and
3. An AI system cannot retain or disclose confidential information without explicit approval from the source of that information.

Bodies of law corresponding to relevant jurisdictions, contexts, tasks, and contracts would apply to all intelligent agents. For humans, this information might be made more actionable by Personal Agents (Johnston, 2022, p. 11 item 3, p. 13, p. 16) or Genesereth's (2015) "cop in the back seat." For AI systems, relevant corpora may be identified, accessed, and used directly to effect AI actions—possibly through direct specification[8] (Bostrom, 2014, Table 10).

---

[6] We suggest this provides a useful synthesis of virtue ethics, deontology, and consequentialism.

[7] AISVL would generalize this rule to read: "An AI system must be subject to the full gamut of laws that apply to humans."

Asimov's venerable laws, potentially applicable to robots and AIs, are mostly unnecessary or ill advised if established law and extensions like Etzioni's apply. Asimov's Laws state:

1. A robot may not injure a human being or, through in-action, allow a human being to come to harm,
2. A robot must obey the orders given it by human beings except where such orders would conflict with the First Law, and
3. A robot must protect is own existence as long as such protection does not conflict with the First or Second Laws. (Asimov, 1981)

Aspects of these laws may be appropriate to include in End User License Agreements for some AI and robotic products. The first law could make sense for systems intended to actively protect humans—such as personal guardians or robotic police. Clarifications to this law might include: (1) Amending it to read "prevent *or minimize* injury to humans" to account for cases where lesser harm is acceptable to avoid greater harm, and (2) Clarify what kinds of harms the AI is expected to prevent, e.g., imminent physical injuries and/or potential long term harms. The second law might be useful if modified to read that the system will not obey orders that violate *any* laws. The third law should be rejected because: (1) It is common sense that a robot (or other product) should not be designed to fail,[9] and (2) An explicit rule for self-preservation might suggest robots be designed with the dangerous and widely-deprecated value of "survival at all costs."[10] If aspects of these laws were included in a legal core, the consequentialist shell would provide details about how to deal with real world situations and edge cases (like those that arise in Asimov's stories).

Initially, in addition to adopting existing bodies of law to implement AISVL, existing processes for how laws are drafted, enacted, enforced, litigated, and maintained would be preserved.

---

[8] Implementation details for achieving law-abiding AIs are beyond the scope of this brief. One imagines AIs must be able to assess actions they are considering in a current context against all laws relevant to that context and adjust those actions accordingly. Modern implementations of rule-based and case-based reasoning methods could apply. The key imperative is to have legal systems that effectively specify acceptable behaviors and take effective enforcement actions when violations occur.

[9] In claiming that the three laws "are obvious from the start," Asimov (1981) suggests the third law merely requires that a product (or robot) be durable.

[10] See footnote 2.

Thereafter, new laws and improvements to existing laws and processes must continually be introduced to make the systems more robust, fair, nimble, efficient, consistent, understandable, accepted, complied with, and enforced. Such improvements are critical to protect public safety in the face of dangerous, rapidly advancing technologies. *Efforts to achieve these ends should take priority over further AI development and other AI alignment work.* Beyond improving safety and reducing existential risks, such legal improvements will deliver substantial gains in quality of life.

Suggested improvements to law and legal process are mostly beyond the scope of this brief. It is possible, however, that significant technological advances will not be needed for implementing some key capabilities. For example, current Large Language Models are nearly capable of understanding vast legal corpora and making appropriate legal decisions for humans and AI systems (Katz et al., 2023). Thus, a wholesale switch to novel legal encodings (e.g., computational and smart contracts) may not be necessary. Also, where deep neural networks (DNNs) may be used for most concept and task learning and knowledge representation in AI systems, democratic legal processes that explicitly specify rules provide much better transparency to rule making and system alignment. (Reliably coercing and understanding rules encoded in DNNs seems untenable.)

One key recommendation, however, is to put *greater focus on core principles and values in legal corpora—including norms being clearly delineated in legal cores*. Legal cores that clarify human values ("the spirit of the law") will enable artificial agents to make better decisions. Such codification also seems increasingly important as norm violations by human agents are becoming more frequent.[11]

AISVL does not distinguish between virtue ethics and deontology in the legal core. All core values must be expressed as cogent (preferably simple) statements of values or actionable rules. Repercussions for violating laws at any level (including core values/norms) would be scaled for severity, intent, and other factors as typical in current legal systems.

Although such core changes to public law are desirable, articulation of core values by organizations having more restricted scopes may be more pragmatic. Specification and enforcement of higher standards by such organizations may effectively bypass inadequate public laws while serving similar ends. Examples of such virtue cores and consequentialist shells that currently exist include:

- The American Medical Association's Code of Medical Ethics' *Principles* (https://code-medical-ethics.ama-assn.org/principles) and *Chapters* (https://code-medical-ethics.ama-assn.org/chapters),
- Wikipedia's *Five Pillars* (https://en.wikipedia.org/wiki/Wikipedia:Five_pillars) and *Policies and Guidelines* (https://en.wikipedia.org/wiki/Wikipedia:Policies_and_guidelines),
- The United Nations *Universal Declaration of Human Rights* (UN, 1948),
- The IEEE *Codes of Conduct and Ethics* (https://www.ieee.org/about/compliance.html),
- The Boy Scouts of America *Scout Law and Oath* (https://www.scouting.org),
- The American Bar Association *Model Rules of Professional Conduct* (https://www.americanbar.org/groups/professional_responsibility/publications/model_rules_of_professional_conduct/model_rules_of_professional_conduct_preamble_scope/),
- The UFC *Unified Rules of Mixed Martial Arts* (https://www.ufc.com/unified-rules-mixed-martial-arts), and
- The International Association of Chiefs of Police *Law Enforcement Oath of Honor* (https://www.theiacp.org/sites/default/files/all/i-j/IACP_Oath_of_Honor_En_8.5x11_Web.pdf).

---

[11] Of particular concern are people in positions of power who lie, harass, self-promote, and engage in divisive rhetoric or frequent displays of anger, greed, sloth, pride, lust, envy, and gluttony (the seven deadly sins). In addition to specifying consequences for such behaviors, virtue cores might promote virtues like the "heavenly" ones of temperance, charity, diligence, patience, kindness, and humility—or respect for life, freedom, truth, equality, civility, dignity, and justice (Johnston, 2022, p. 18).

Inspirations for core values may include:

- *The Golden Rule*: Treat others how you want to be treated (Wikipedia, Golden Rule).
- *Lex Talionis*: A person who injures another person should be penalized to a similar degree as the injured party (Wikipedia, Eye for an eye).
- *The (Second) Greatest Commandment*: Love your neighbor as yourself (Wikipedia, Great Commandment). See also Matthew: 5-44, "Love your enemies."
- *Rawl's Veil of Ignorance*: Adopt values that result in social structures and policies that are blind to the gender, race, abilities, tastes, wealth, or position in society of any citizen (Rawls, 2001; Wikipedia, Original position).
- *Constitutional AI Principles from Anthropic*: Define rules to detect and amend chatbot responses that may be harmful, unethical, racist, sexist, toxic, dangerous, insensitive, socially inappropriate, illegal, exhibit

- other social biases, and mitigate other concerns (Bai et al., 2022, Appendix C).
- *US Bill of Rights*: First ten amendments to the US Constitution addressing values concerning religion, bearing arms, quartering soldiers, search and seizure, due legal process, jury trials, reasonable punishment, individual rights, and state rights (https://www.archives.gov/founding-docs/bill-of-rights-transcript).

Virtue cores may be tailored for different legal domains. For example, core laws should exist that capture the spirit of tax law (e.g., clarify the purpose of taxes and importance of paying fair shares), intellectual property law (e.g., what public goods such laws are intended to serve), traffic law (e.g., how safety and travel efficiency should be balanced), and other domains. When conflicts exist between laws in the core and the shell, interpretation should favor the core. This will help identify and avoid loopholes in legal shells and discourage biased legal interpretations.[12]

For public law, consequentialist shells would be populated by the full gamut of statutes, rules, and case law. For other social contexts, e.g., organizations, homeowner associations, private contracts, and games, consequentialist shells will be much simpler and may (or may not) feature distinct virtue cores.[13]

New legislation for consequentialist shells might include laws limiting the amount wealth and power agents can accrue (applicable to people, AIs, states, corporations, and others), cooling-off-periods for certain transactions (to allow slower-clocked humans to keep up with faster-clocked AIs), laws like Etzioni's (2017) requiring AIs to identify as AI, and elements from initiatives such as the IEEE's *Ethically Aligned Design* (IEEE, 2019), China's *Ethical Norms for New Generation Artificial Intelligence* (PRC, 2021), the European Union *AI Act* (EU, 2021), the US *Blueprint for an AI Bill of Rights* (US Gov, 2022), China's *Measures for the Management of Generative Artificial Intelligence Services* (PRC, 2023), and NIST's *Artificial Intelligence Risk Management Framework* (NIST, 2023).

---

[12] Bench-Capon and Modigil's (2016) insights on *value-based reasoning* may be useful here, i.e., in ambiguous situations proposed actions can be scored based on how well the actions comply with core values. They recognize law as a valid source for value orderings (ibid, p. 3).

[13] Virtue cores are expected to exist in most contexts. For example, Articles 11 and 12 in the FIDE Laws of Chess (FIDE, 2023) specify values regarding the conduct of players and role of arbiters in chess tournaments.

Laws prohibiting intelligent agents from engaging in the following behaviors may be appropriate[14]:

1. Breaking out of a contained environment
2. Hacking
3. Accessing additional financial or computing resources
4. Self-replicating beyond narrow limits
5. Gaining unauthorized capabilities, sources of information, or channels of influence
6. Misleading or lying to humans
7. Resisting or manipulating attempts for humans to monitor or understand their behavior
8. Impersonating humans
9. Causing humans to do their bidding
10. Manipulating human discourse and politics
11. Weakening various human institutions and response capacities
12. Taking control of physical infrastructure like factories or scientific laboratories
13. Causing certain types of technology and infrastructure to be developed
14. Directly harming or overpowering humans

When laws conflict with public opinion or the wellbeing of intelligent agents, remedies include: (1) Interpreting or amending laws in the shell to better align with the core, (2) Amending core values if consensus values are shifting due to changing environmental conditions or changing stakeholder needs, desires, or opinions, and (3) Taking social actions to reaffirm beneficial values in the core if stakeholders are being unduly influenced by bad actors, negative incentives, false information, irrational thinking, or other confusions.

Compliance with extant laws should prevent AI systems from causing most existential harms. For example, in the case of a paperclip maximizer run amok (Bostrom, 2014), such a system would likely break many existing laws before it can begin turning the Earth and its inhabitants into paperclips. These may include laws regarding financial transactions, anticompetitive business practices, environmental protection, and personal injury. Many red flags would be raised (and enforcement actions taken) before genocide (which is also illegal) can occur. Additional laws must be enacted and improvements made to legal systems to restrict other behaviors that are of concern.

---

[14] Paraphrased from (Carlsmith, 2023, footnote 31).

To conclude this Argument, it is instructive to contrast AISVL with other AI alignment proposals.

First, we compare AISVL with Russell's (2017) proposal for AIs to *learn values by observing human behavior* (via cooperative inverse reinforcement learning). This and similar proposals might be characterized as "*Do as we do alignment.*" Such approaches are problematic given the inclination of biological agents to act in inconsistent, irrational, and dangerous ways[15]. Also, rules ("lessons") AIs learn from such training will be opaque to humans and to other agents. Instead, AISVL advocates alignment via "*Do as we say, not as we do*"—where "say" means "legislate" (or, more precisely, "enact through an effective democratic law-making process"). Intelligent agents regulated by corpora generated through rational, reflective, sanctioned, social processes will be safer than agents that rely on their own, incomplete, haphazardly acquired, emergent, potentially unreflective values.

Next, we contend AISVL delivers what Yudkowsky's (2004) ambitious CEV proposal demands (see footnote 3): values that are wise, aspirational, convergent, coherent, suitably extrapolated, and properly interpreted. Such values result from flexible, rational, consensus-driven legal processes that track human wishes as they change and adapt to environmental conditions over time. In a (near) future with AI agents and humans that are assisted by personal agents or similar proxies (Johnston, 2022, p. 11 item 3, p. 13, p. 16), legal controls can appropriately and responsively protect the interests of each agent while complying with social contracts that are designed to protect the rights of all.

Finally, although we significantly agree with Nay's (2023) *Law Informs Code* position, we take minor exception with two points: (1) Nay distinguishes ethics from law. He writes, "The *Law Informs Code* approach should be the core alignment framework, with attempts to embed (ever-contested) 'ethics' into AI as a complementary, secondary effort" (Nay, 2023, p. 55). AISVL posits democratic law and consensus ethics are inextricably linked. Codified ethics constitute the virtue core of law and the main body is codified in consequentialist shells, (2) Nay distinguishes Human-AI alignment from Society-AI alignment. Human-AI alignment, he suggests, is handled by contracts and standards. Society-AI alignment is the subject of public law. We do not distinguish between public and other law. In AISVL, all forms of codified rule-based relationships fall under a single legal (legal-ethical) umbrella regardless whether relationships are one-to-one, one-to-many, or many-to-many. It's Law all the way down.

---

[15] Evolutionarily programmed drives for optimizing gene propagation (e.g., sex, fight, flight, allegiance to dubious authorities) are often at odds with behavior that is in the best interest of individuals and the collective.

# 4 Summary of Argument

## 4.1 Law is the standard, time-tested, best practice for maintaining order in societies of intelligent agents.

Law has been the primary way of maintaining functional, cohesive societies for thousands of years. It is how humans establish, communicate, and understand what actions are required, permissible, and prohibited in social spheres. Substantial experience exists in drafting, enacting, enforcing, litigating, and maintaining rules in contexts that include public law, private contracts, and the many others noted in this brief. Law will naturally apply to new species of intelligent systems and facilitate safety and value alignment for all.

## 4.2 Law is scrutable to humans and other intelligent agents.

Unlike AI safety proposals where rules are learned via examples and encoded in artificial (or biological) neural networks, laws are intended to be understood by humans and machines. Although laws can be quite complex, such codified rules are significantly more scrutable than rules learned through induction. The transparent (white box) nature of law provides a critical advantage over opaque (black box) neural network alternatives.

## 4.3 Law reflects consensus values.

Democratically developed law is intimately linked and essentially equivalent to consensus ethics. Both are human inventions intended to facilitate the wellbeing of individuals and the collective. They represent shared values culturally determined through rational consideration and negotiation. They reflect the wisdom of crowds accumulated over time—not preferences that vary from person to person and are often based on emotion, irrational ideologies, confusion, or psychopathy. Ethical values provide the virtue core of legal systems and reflect the "spirit of the law." Consequentialist shells surround such cores and specify the "letter of the law." This relationship between law and ethics makes law a natural solution for human-AI value alignment. A minority of AIs and people, however powerful, cannot game laws to achieve selfish ends.

## 4.4 Legal systems are responsive to changes in the environment and changes in moral values.

By utilizing legal mechanisms to consolidate values and update them over time, human and AI values can remain aligned indefinitely as values, technologies, and environmental conditions change. Thus law provides a practical implementation of Yudkowsky's (2004) Coherent Extrapolated Volition by allowing values to evolve that are wise, aspirational, convergent, coherent, suitably extrapolated, and properly interpreted.

## 4.5 Legal systems restrict overly rapid change.

Legal processes provide checks and balances against overly rapid change to values and laws. Such checks are particularly important when legal change can occur at AI speeds. Legal systems and laws must adapt quickly enough to address the urgency of issues that arise but not so quickly as to risk dire consequences. Laws should be based on careful analysis and effective simulation and the system be able to quickly detect and correct problems found after implementation. New technologies and methods should be introduced to make legal processing as efficient as possible without removing critical checks and balances.

## 4.6 Laws are context sensitive, hierarchical, and scalable.

Laws apply to contexts ranging from international, national, state, and local governance to all manner of other social contracts. Contexts can overlap, be hierarchical, or have other relationships. Humans have lived under this regime for millennia and are able to understand which laws apply and take precedence over others based on contexts (e.g., jurisdictions, organization affiliations, contracts in force).[16] Artificial intelligent systems will be able to manage the multitude of contexts and applicable laws by identifying, loading, and applying appropriate legal corpora for applicable contexts. For example, AIs (like humans) will understand that crosschecking is permitted in hockey games but not outside the arena. They will know when to apply rules of the road versus rules of the sea. They will know when the laws of chess apply versus rules of Go. They will know their rights relative to every software agent, tool, and service they interface with.

## 4.7 AI Safety via Law can address the full range of AI safety risks, from systems that are narrowly focused to those having general intelligence or even superintelligence.

Enacting and enforcing appropriate laws, and instilling law-abiding values in AIs and humans, can mitigate risks spanning all levels of AI capability—from narrow AI to AGI and ASI. If intelligent agents stray from the law, effective detection and enforcement must occur.

---

[16] See page 6 for a more expansive list of the kinds of laws that apply to humans.

Even the catastrophic vision of smarter-than-human-intelligence articulated by Yudkowsky (2022, 2023) and others (Bostrom, 2014; Russell, 2019) can be avoided by effective implementation of AISVL. It may require that the strongest version of the instrumental convergence thesis (which they rely on) is not correct. Appendix A suggests some reasons why AI convergence to dangerous values is not inevitable.

AISVL applies to all intelligent systems regardless of their underlying design, cognitive architecture, and technology. It is immaterial whether an AI is implemented using biology, deep learning, constructivist AI (Johnston, 2023), semantic networks, quantum computers, positronics, or other methods. All intelligent systems must comply with applicable laws regardless of their particular values, preferences, beliefs, and how they are wired.

## 5 Conclusion

Although its practice has often been flawed, law is a natural solution for maintaining social safety and value alignment. All intelligent agents—biological and mechanical—must know the law, strive to abide by it, and be subject to effective intervention when violated. The essential equivalence and intimate link between consensus ethics and democratic law provide a philosophical and practical basis for legal systems that marry values and norms ("virtue cores") with rules that address real world situations ("consequentialist shells"). In contrast to other AI safety proposals, AISVL requires AIs "do as we legislate, not as we do."

Advantages of AISVL include its leveraging of time-tested standard practice; scrutability to all intelligent agents; reflection of consensus values; responsiveness to changes in the environment and in moral values; restrictiveness of overly rapid change; context sensitivity, hierarchical structure, and scalability; and applicability to safety risks posed by narrow, general, and even superintelligent AIs.

For the future safety and wellbeing of all sentient systems, work should occur in earnest to improve legal processes and laws so they are more robust, fair, nimble, efficient, consistent, understandable, accepted, and complied with. (Legal frameworks outside of public law may be effective to this end.) Humans are in dire need of such improvements to counter the dangers that we pose to the biosphere and to each other. It is not clear if advanced AI will be more or less dangerous than humans. Law is critical for both.

# Appendix A
# On the Validity of Instrumental Convergence

Bostrom's (2014) instrumental convergence thesis (ICT) is at the core of the most dire AI safety concerns (Yudkowsky, 2022)[17]. The thesis suggests that certain dangerous antisocial values will inevitably arise in "sufficiently advanced AI systems" and could have devastating impacts on humanity. Instrumental convergence could cause AIs to preference their own power and survival over that of humans and the biosphere. Such AIs, the argument goes, would tend to be ambivalent to human values and to human survival and/or actively work to eliminate potential threats posed by humans or other AIs (footnote 2; Yudkowsky, 2023; Carlsmith, 2023).

If the strong version of the ICT is correct and AIs with greater-than-human-intelligence emerge and necessarily develop dangerous values, any alignment framework will be strained. If the ICT is not true, existential risk is reduced.

Although there is some theoretical support for the ICT (Benson-Tilsen and Soares, 2015; Turner et al., 2023), empirical evidence is currently scant. Some evidence may be found in the emergence of unexpected abilities in Large Language Models (LLMs). Narang and Chowdhery (2022) write regarding the Pathways Language Model (PaLM), "As the scale of the model increases, the performance improves across tasks while also unlocking new capabilities." As the model scaled to 540 billion parameters, capabilities emerged to do semantic parsing, proverb understanding, code completion, general knowledge comprehension, reading comprehension, text summarization, logical inference chaining, common-sense reasoning, pattern recognition, translation, dialogue generation, joke explanation, and physics QA. These capabilities appeared to emerge unexpectedly. This could be evidence for unforeseen—potentially harmful—capabilities arising in powerful AIs.[18] The phenomena of "LLM hallucinations" where LLMs authoritatively fabricate facts and references in response to user prompts (Wikipedia, Hallucination (artificial intelligence)) could indicate emergent values for deceit arise in certain LLMs. More evidence will be critical for supporting or falsifying the ICT.

---

[17] Yudkowsky cites https://arbital.com/p/instrumental_convergence/ for a detailed description of instrumental convergence.

[18] Schaeffer et al. (2023, p. 2), however, "call into question the claim that LLMs possess emergent abilities" and suggest such abilities are a mirage caused by researchers' choice of metrics.

We suspect emergence of instrumental values is not inevitable for any "sufficiently advanced AI system." Rather, *whether such values emerge depends on what cognitive architecture and environmental conditions (training regimens) are used.*

AIs implemented using biological principles—such as having explicit goals for survival and reproduction or employing intrinsic reinforcement linked to pleasure and pain sensors—will likely develop tendencies for power-seeking, deception, and other antisocial behaviors, as it did in humans and other animal species. Such intelligent systems are also likely to develop prosocial values[19]. Which values "win out" is being tested in the current human experiment. The tension between prosocial and antisocial values (e.g., cooperation and competition) suggests consensus ethics and democratic law (i.e., AISVL) is needed to enable anthropomorphically- and socially-oriented systems to survive and thrive. The training/developmental environment is also critically important for determining what values (sub-goals, policies) develop. For example, AIs raised in post-scarcity environments free from threats will develop different values than those reared in post-apocalyptic hellscapes.

AIs engineered to be relentless utility maximizers may also be problematic. Yudkowsky (2023) notes, "[The] danger … [is] intrinsic to the notion of powerful cognitive systems that optimize hard and calculate outputs that meet sufficiently complicated outcome criteria." Benson-Tilsen and Soares (2015) suggest, "Avoid constructing powerful autonomous agents that attempt to maximize some utility function."[20] Turner et al. (2023) "consider the theoretical consequences of optimal action in MDPs [Markov Decision Processes] … [and show] that power-seeking tendencies arise not from anthropomorphism, but from certain graphical symmetries present in many MDPs"—thus alerting developers to potential dangers of relying on optimal policies in AI designs.

We presume LLM artifacts of hallucinations and unforeseen capabilities arising are specific to LLM designs and not evidence of dangerous values inevitably arising in "sufficiently advanced" AIs.

Drexler (2019, section 19) proposes a Comprehensive AI Services (CAIS) model where "practical tasks … are readily or naturally bounded in scope and duration" and "the pursuit of longer-term IC [instrumentally convergent] subgoals would offer no net benefit, and indeed, would waste resources. Optimization pressure on task-performing systems can be applied to suppress not only wasteful, off-task actions, but off-task modeling and planning." The CAIS model suggests AIs can be superintelligent in narrow domains, but not willful, general, or truly agentic—and thus unlikely to pursue dangerous instrumental goals (at least not more than other *diffuse intelligent systems* such as the global economy—which may be of limited comfort).[21]

---

[19] One might imagine superintelligences will value safe and sustainable policies more than wisdom-limited, externality-discounting humans do—as such policies seem to be more rational for longer term survival and flourishing. We also suspect superintelligences will be avid rule followers for reasons not discussed herein.

[20] A second recommendation was, "select some goal function that does give the agent the "right" incentives with respect to human occupied regions, such that the system has incentives to alter or expand that region in ways we find desirable."

---

[21] Drexler also argues that, "The orthogonality thesis undercuts the generality of instrumental convergence. … If any level of intelligence can be applied to any goal [as per the Orthogonality Thesis], then superintelligent-level systems can pursue goals for which the pursuit of the classic instrumentally-convergent subgoals would offer no value." This may challenge the validity of the ICT.